\documentclass[10pt,conference]{IEEEtran} 
\IEEEoverridecommandlockouts
\usepackage{cite}
\usepackage{amsmath,amssymb,amsfonts}
\usepackage{algorithmic}
\usepackage{graphicx}
\usepackage{textcomp}
\usepackage{xcolor}
\usepackage{subcaption}
\usepackage{booktabs} 
\usepackage[T1]{fontenc}
\usepackage[utf8]{inputenc}
\def\BibTeX{{\rm B\kern-.05em{\sc i\kern-.025em b}\kern-.08em
    T\kern-.1667em\lower.7ex\hbox{E}\kern-.125emX}}
\begin{document}

\title{Predictive Auxiliary Learning for Belief-based Multi-Agent Systems
}

\author{\IEEEauthorblockN{Qinwei Huang}
\IEEEauthorblockA{\textit{Electrical Engineering and Computer Sciences} \\
\textit{Syracuse University}\\
Syracuse, USA \\
qhuang18@syr.edu}
\and
\IEEEauthorblockN{Stefan Wang}
\IEEEauthorblockA{\textit{Computer Science} \\
\textit{University of Rochester}\\
Rochester, USA \\
swang170@u.rochester.edu}
\and
\IEEEauthorblockN{Simon Khan}
\IEEEauthorblockA{\textit{Air Force Research Laboratory} \\
%\textit{name of organization (of Aff.)}\\
%City, Country \\
simon.khan@us.af.mil}
\and
\IEEEauthorblockN{Garrett E. Katz}
\IEEEauthorblockA{\textit{Electrical Engineering and Computer Sciences} \\
\textit{Syracuse University}\\
Syracuse, USA \\
gkatz01@syr.edu}
\and
\IEEEauthorblockN{Qinru Qiu}
\IEEEauthorblockA{\textit{Electrical Engineering and Computer Sciences} \\
\textit{Syracuse University}\\
Syracuse, USA \\
qiqiu@syr.edu}
}
\maketitle

\begin{abstract}
The performance of multi-agent reinforcement learning (MARL) in partially observable environments heavily relies on the effective aggregation of information from both observations and communications, as well as feedback signals like rewards. While most state-of-the-art multi-agent systems (MAS) primarily leverage rewards as the sole feedback for policy training, our research reveals that training agents to perform auxiliary predictive tasks  can substantially enhance system performance. In this paper, we introduce BElief-based Predictive Auxiliary Learning (BEPAL), a novel approach that leverages auxiliary training objectives to provide additional information, aiding in policy learning. BEPAL follows the centralized training with decentralized execution (CTDE) paradigm. Each agent using BEPAL trains a belief model that predicts unobservable state information, such as other agents’ rewards and motion directions, alongside its policy model. By enriching hidden states to represent critical environmental information that does not directly contribute to reward maximization, this concurrent auxiliary learning stabilizes the MARL process and improves its performance. We validate the effectiveness of BEPAL using two multi-agent cooperative games: the predator-prey game and the Google Research Football (GRF). In these scenarios, BEPAL outperforms existing methods, achieving an average 16\% improvement in performance metrics and more stable convergence. 
\end{abstract}


\section{Introduction}
Multi-Agent Systems (MAS) are extensively employed for tasks that require coordinated efforts among multiple autonomous agents, such as collaborative robotics and complex game environments. In MAS, the presence of multiple agents introduces greater uncertainty, increases computational demands for communication, and complicates the achievement of game-theoretic equilibria. To mitigate these challenges, Multi-Agent Reinforcement Learning (MARL) is utilized, enabling agents to learn to make optimal decisions through interaction with their environment and fellow agents.

Centralized Training with Decentralized Execution (CTDE) is a widely adopted strategy in MARL, designed to address the non-stationarity \cite{kraemer2016multi,oliehoek2008optimal} that arises when multiple agents learn and adapt simultaneously. CTDE uses a centralized value function during training, which is later decomposed into individual utility functions for decentralized execution, aligning agents' actions with both individual and collective objectives. Moreover, recent advancements \cite{oroojlooy2023review,simoes2020multi,iqbal2019actor} have incorporated communication systems into Centralized Critic with Decentralized Actor models to better manage challenges in partially observable environments. Previous works on communication-based Centralized Critic with Decentralized Actor  \cite{singh2018learning,niu2021multi,das2019tarmac} have demonstrated that integrating a communication system can significantly improve MARL performance. While these methods generally improve cooperation among agents, achieving high performance in complex tasks remains a challenge.  

A DRL agent maps environment states to actions for maximum expected reward. The pursuit of enhanced state representation with richer information has recently become a key focus in addressing multi-agent challenges. For example, to obtain enriched state information, \cite{ayala2022kraf} integrates knowledge graphs with agent observations, \cite{pu2022attention} leverages attention mechanisms to enhance state by focusing on critical communication messages, and \cite{lin2024enhancing} combines predicted future events with current observation. However, these works adopt centralized control. They assume that the centralized controller has global observability and maps the observation plus auxiliary information (i.e., knowledge graph or prediction) directly to action. If their state enhancement method has trainable parameters, they are trained directly to maximize the rewards.

Many decentralized multi-agent games are partially observable systems. Agents can only observe their local environment and depend on communication to achieve global awareness. Additionally, these games are often non-Markovian from a single agent's perspective. The expected reward distribution depends not only on the target agent’s current state but also on the states of other agents and hidden environment variables, which can be influenced by the target agent’s past actions. Existing approaches, such as \cite{sukhbaatar2016learning, singh2018learning, das2019tarmac}, use Long short-term memory (LSTM) models to integrate an agent’s historical observations and received messages into a hidden state. This hidden state needs to be comprehensive, capturing both current and past information about the environment. Decision-making based on this hidden state is considered to be Markovian. We refer to this hidden state as the \emph{belief state} as it represents the agent's understanding of the environment. The belief state representation is trained through reinforcement learning using reward feedback.

Prior work has explored improving  belief state representation by training agents to estimate the current locations of other agents, demonstrating enhanced policy performance \cite{luo2023multi}. However, this technique is not scalable to larger and more complex environment because it models the environments as a grid-based system and applies CNN to encode the observations. Moreover, to further enhance the belief representation, the agent needs to do more than estimating the location of other agents. 

This work presents BElief-based Predictive Auxiliary Learning (BEPAL), a novel MARL technique that enhances agents' belief state representation by auxiliary predictive learning. BEPAL trains the agent to perform auxiliary tasks, such as predicting critical unobservable or partially observable information like rewards and the locations of other agents, to improve the performance of MARL. BEPAL supplements each agent with a belief decoder and a set of auxiliary objective functions, which train the decoder to transform the agent’s hidden state into a neuro-symbolic representation of its understanding of the global environment. This representation may contain predictive information of other agents' locations, moving directions, and future rewards. While they may not directly affect the reward of the target agent, they influence the outcome of collaboration activities, and, consequently, the overall system reward. Our experimental results demonstrate that enriching the belief state representation by training agents to perform these auxiliary predictive tasks significantly improves performance in multi-agent distributed systems where agents make local decisions.

Additionally, we employ a multi-head graph attention neural network (GAT) \cite{velivckovic2017graph} as the observation encoder to handle complex observations in non-grid based environment. The complexity of this encoder is proportional to the size of the agent's observations and independent to the environment's size. Unlike convolutional neural networks (CNNs), which operate on Euclidean vector space, the GAT based encoder accommodates arbitrary precision in object locations and non-Euclidean inputs.  With these enhancements, BEPAL not only surpasses previous models in agent performance but is also adaptable to new environments that were unsupported by some of the earlier models. 

\section{Related Works}

In multi-agent settings, each agent can perceive other agents as part of their state observations, resulting in a dynamic and non-stationary environment \cite{ye2015multi, xu2018multi}. Centralized Training with Decentralized Execution (CTDE) effectively addresses these challenges by maintaining strong generalization abilities and resolving non-stationarity better than centralized control methods like QMIX \cite{rashid2020monotonic}. A key advancement in this area is the use of a centralized critic in actor-critic algorithms, which facilitates decentralized policy learning, as demonstrated by Multi-Agent Deep Deterministic Policy Gradient (MADDPG) \cite{lowe2017multi}. Despite its strengths, MADDPG struggles with inadequate experience collection, impeding performance and complicating convergence. COMA \cite{foerster2018counterfactual} further improves centralized critic with counterfactual baselines to enhance credit assignment and optimize decentralized policies in cooperative multi-agent systems.

While these approaches mark critical improvements in MARL, the introduction of communication systems among agents has proven essential for further enhancing performance. However, the complexity of applying joint value functions increases with the number of agents, making centralized training more challenging \cite{yao2021smix, yang2020q, zhou2020learning, su2021value}. Innovations like Reinforced Inter-Agent Learning (RIAL) and Differentiable Inter-Agent Learning (DIAL) \cite{foerster2016learning} introduced message-passing frameworks based on local observations and actions. CommNet \cite{sukhbaatar2016learning} integrated a centralized communication channel, while IC3Net \cite{singh2018learning} added a communication gate to optimize timing. However, these models often overlook the content of messages, relying on simple averaging and equal weighting, which may not capture the full dynamics of the environment. TarMAC \cite{das2019tarmac} and MAGIC \cite{niu2021multi} address this by employing attention mechanisms to selectively weight and direct message passing, improving the relevance of shared information. PR2 \cite{wen2019probabilistic} approximates opponents’ conditional policies using variational Bayes methods to iteratively update the agent's own strategies.

Despite these advancements, many MARL approaches overlook key aspects of team dynamics and environmental context, such as the varying positions and rewards of other agents. In DRL, auxiliary objectives have proven effective for enhancing state representation by predicting rewards or future states \cite{jaderberg2016reinforcement, shelhamer2016loss, oord2018representation, wayne2018unsupervised}. In MARL, however, it is essential to integrate information about other agents' states to improve the overall understanding of the environment. BAMS \cite{luo2023multi} uses CNNs on grid maps to provide additional feedback, though it is limited to grid-based environments and incurs additional computational expense.

Our approach differs by extracting key information within MAS for auxiliary prediction objectives to enrich the agents' belief states. By integrating predictions of rewards from other agents, we enable each agent to build a more accurate understanding of the overall multi-agent system, leading to improved cooperation and performance.

\section{Preliminaries}
\subsection{Dec-Partially Observable Markov Decision Processes}

We investigate a cooperative multi-agent game within the framework of the Decentralized Partially Observable Markov Decision Process (Dec-POMDP), defined as the tuple $\langle N, S, P, \mathcal{R}, \mathcal{O}, \mathcal{A}, Z, \gamma\rangle$. Here, $N$ denotes the number of agents; $S$ is the finite state space; $P(s^{\prime} \mid s, a) : S \times \mathcal{A} \times S \rightarrow [0,1]$ represents the state transition probabilities; $\mathcal{A} = [\mathbf{A}_{1} \ldots \mathbf{A}_{N}]$ constitutes the finite set of actions, with $\mathbf{A}_{i}$ indicating the local actions $\mathbf{a}_{i}$ available to agent $i$; $\mathcal{O} = [\mathbf{O}_{1} \ldots \mathbf{O}_{N}]$ encapsulates the finite set of observations governed by the observation function $Z: S \times \mathcal{A} \rightarrow \mathcal{O}$; $\mathcal{R}: S \times \mathcal{A} \rightarrow \mathbb{R}^N$ is the reward function; and the constant $\gamma \in (0,1]$ is a discount factor. In each time step $t$, agent $i$ selects action $\mathbf{a}_{i}^{t}$, and receives reward ${r}_{i}^{t}$ and observation ${o}_{i}^{t}$. Agent $i$ aims to maximize its discounted reward $R_{i} = \sum_{t=0}^{T}\gamma^{t} r^{t}$.

\subsection{Multi-head Graph Attention Networks}
\label{mhgat}
Graph Attention Networks (GATs) leverage a self-attention mechanism to refine node representations within a graph. Each node $i$ evaluates the importance of its neighboring nodes $j$, where $j\in{\mathcal{N}_i}$ and ${\mathcal{N}}_i$ is the set of one-hop neighbors of $i$. The attention coefficient $\delta_{ij}$ is calculated as follows: 
\begin{equation}
w_{ij} =LeakyReLU(a^T [W f_i \Vert W f_j])
\end{equation}
\begin{equation}
\delta_{ij} = \frac{e^{w_{ij}}}{\sum_{k\in{\mathcal{N}_i}} e^{w_{ik}}} \ \ \ for\ j=1,2,\dots,|{\mathcal{N}_i}|
\end{equation}

where $[\cdot \Vert\cdot]$ denotes concatenation, $f_i$ and $f_j$ are features of nodes $i$ and $j$, $W$ is a trainable weight matrix, and $a$ is a trainable weight vector.

Using the attention coefficient, node $i$ updates its feature vector by aggregating information from its neighbors:
\begin{equation}
    \vec{f_i}= LeakyReLU\left(\sum_{j\in \mathcal{N}_i\bigcup{\{i\}}}{\delta_{ij}W{f_j}}\right)
    \label{gat1}
\end{equation}
To encapsulate more complex relationships among nodes, a multi-head GAT configuration is employed. Each attention head computes its own attention coefficient $\delta_{ij}^k$ and produces an updated feature vector $\vec{f_i}^k$, where $k$ denotes the  head index. 

The GAT layer updates a node's representation using the features of its one-hop neighbors. To incorporate the features from N-hop neighbors, an N-layer of GAT is required.  

\section{Method}
\label{method}

\begin{figure*}
\vspace{1pt}
  \centering
  \centerline{\includegraphics[scale=0.5]{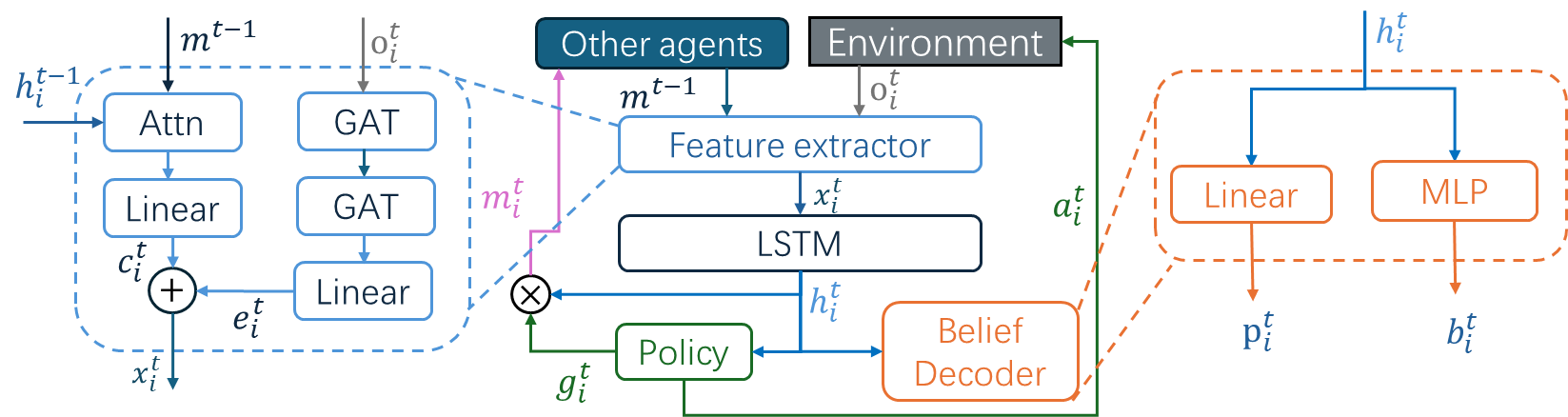}}
  \caption{The BEPAL-MAS architecture.}
  \label{framework}
  \vspace{-7pt}
\end{figure*}

Our target system comprises $N$ agents, each equipped with broadcasting-based communication capabilities. From each agent's perspective the environment is modeled as a POMDP. At each time step $t$, an agent $i$ receives a local observation $o_i^t$ and inter-agent messages $m_{ij}^{t-1}$ from other agents $1\leq{j}\leq{N}$. Using this information, the agent updates its hidden state $h_i^t$. A policy $\pi_i$ then maps the hidden state $h_i^t$ to a probability distribution over the action space. Each action consists of two components: a communication action that determines whether message broadcasting is gated or allowed, and a movement action that dictates how the agent plays the game. The actual action $a_i^t$ is sampled according to $a_i^t \sim \pi_i$.

In this work, we focus on homogeneous agents and  adopt a model sharing approach, where the same model is used by different agents with their local observations $o_i$. The model enables the agents to maintain their local hidden states $h_i$ and select local actions $a_i$. A shared model typically can be trained faster than the agent specific model because $N$ times more training data can be collected from each game. 
%parameter-sharing framework in which a single large network, consisting of shared LSTM-based components, is used to control each agent independently. Each agent processes its own local observations and messages, and while only one LSTM output generates the action for the respective agent, the shared parameters ensure efficiency and scalability.
%The policy $\pi_i$ is recurrent with hidden state $h_i^t$ conditioned on both $o_i^t$ and $m_{ij}^{t-1}$.  

Each agent also uses the hidden state $h_i^t$ to update its beliefs $b_i^t$ and $p_i^t$ about other agents' motion and rewards, respectively.  Maximizing the accuracy of these beliefs serves a set of auxiliary training objectives. The subsequent subsections will detail the operational flow of the framework, describe the feature extraction models for the received input, elaborate on the functionalities of the belief decoder module, and delineate the loss functions used to train the system.

The architecture of the BEPAL model is depicted in Fig \ref{framework}. It consists of a feature extractor, an LSTM, a policy network and a belief decoder network. 

\subsection{Feature Extractors}

For agent $i$ at timestamp $t$, the inputs, which consisting of local observations $o_i^t$ and inter-agent messages $m_{ij}^{t-1}$, are processed by the feature extractor. The observation $o_i^t$ is represented as a graph $G(Vertex, Edge)$, where each vertex represents an object observed by the agent except  vertex 0, which presents the agent itself. We consider only the relationship between the observed objects and the agent, hence the graph has a star connection, $Edge=\{({edge}_0, {edge}_i)|\forall{j},{0<j\leq{N}\}}.$ Each vertex in the graph is associated with a feature vector. 

%The node features are designed specifically for the applications. 
%\textcolor{red}{Specifically, $o_i^t$ includes a node feature matrix, which represents the objects observed by the agent, and an edge matrix, which indicates the connections between the agent and the observed objects.} 

The feature extractor processes the agent's observations using a two-layer Graph Attention Network (GAT), trained to assign higher attention weights to objects containing critical information, such as goals or enemies, to enable more informative feature extraction. The first layer employs 3 attention heads while the second layer has a single attention head. We select only the GAT output for node 0, $\vec{f_0}$, as it represents the feature of agent $i$ updated with aggregated information from its observations. $\vec{f_0}$ is further processed by a fully connected layer, and the output, $e_i^t$, serves as the encoded observation of agent $i$. 

In addition to local observations, communication messages exchanged between agents play a crucial role in decision-making. These messages encapsulate historical information gathered by each agent and are shared with their teammates. This approach is inspired by frameworks utilized by TarMAC and IC3Net, where the hidden state held by local LSTM also serves as the communication message. Not all received messages are equally important to an agent’s decision-making, highlighting the importance of efficient information extraction. To address this, an attention mechanism, similar to the GAT used for observations, is employed for effective message aggregation.
The query, key and value ($q_i^t$, $k_j^t$, $u_j^t$) of the attention model are generated from the agent's hidden state and received messages by passing through a Linear layer respectively, $q_i^t=W_qh_i^t$, $k_j^t=W_km_j^{t-1}$, $u_j^t=W_um_j^{t-1}$, where $W_q$, $W_k$ and $W_u$ are trainable weight matrices. The aggregated message $c_i^t$ is then computed following Equations \ref{attn} and \ref{add}:
\begin{equation}
\boldsymbol{\alpha}_i^t=\textrm{SoftMax}\left[\frac{\left(q_i^{t^T} k_1^t\right)}{\sqrt{\left(d_k\right)}}, \ldots ,\frac{\left(q_i^{t^T} k_j^t\right)}{\sqrt{\left(d_k\right)}}, \ldots ,\frac{\left(q_i^{t^T} k_N^t\right)}{\sqrt{\left(d_k\right)}}\right]
\label{attn}
\end{equation}
\begin{equation}
c_i^t=\sum_{j=1}^N \alpha_{ij}^t u_j^t
\label{add}
\end{equation}

\subsection{Hidden State Update and Decision Making}

The aggregated observation ($e_i^t$) and message ($c_i^t$) are summed up as $x_i^t$ and fed into an LSTM network. This LSTM enables the agent to retain historical information, making it easier to learn long-term goals and facilitating better policy learning by approximating the hidden state of the latent Markovian Process of the environment. The agent’s hidden state $h_i^t$ and cell state $s_i^{t+1}$ are updated using Equation \ref{lstm}.
\begin{equation}
h_i^{t+1}, s_i^{t+1} = lstm(e_i^t+c_i^t, h_i^t, s_i^t)
\label{lstm}
\end{equation}
Based on the hidden state, the agent chooses actions using a trained policy network within the actor-critic framework. The actor network, denoted as $actor()$, consists of a fully connected linear layer with LogSoftmax activation function. It maps the hidden state to a vector of action probabilities, $\pi(a|h)$, also known as policy. The action taken by agent $i$ at timestamp $t$, denoted as $a_i^t$, is sampled from the policy distribution $a_i \sim \pi(a_i^t|h_i^t)$. 

At each step, an agent must make two decisions, (1) to select a movement action to advance the game objectives, and (2) to decide whether to broadcast its hidden state. These actions are denoted as $a_i^t$ and $g_i^t$ respectively. The corresponding policies are denoted as $\pi_a$ and $\pi_g$. 
The outgoing message $m_i^t$ for agent $i$ at timestamp $t$ is computed as the element-wise product of hidden state and gate action: $m_i^t = h_i^t \odot g_i^t$. 

The critic model, denoted as $critic()$, is a single-layer fully connected network. It gives agent $i$'s local estimation, $v_i^t$, of the total discounted future rewards. 
\begin{equation}
    a_i^t, g_i^t \sim actor(h_i^t),
    \quad
    v_i^t = critic(h_i^t)
    \label{actor}
\end{equation}

 \subsection{Auxiliary Predictive Tasks}

In addition to the policy network, the hidden state $h_i^t$ is also used as the input for a belief decoder, which performs two auxiliary tasks: reward prediction and motion prediction. The reward prediction is handled by a single layer fully-connected linear network $p_i^t=P_\theta(h_i^t)$. The output, $p_i^t\in{R^N}$, is a vector of expected rewards. The $j$th entry in ${p_i^t}$ gives agent $i$'s estimation of the discounted future reward of agent $j$. The motion prediction uses a 2-layer fully connected linear network, $b_i^t=B_\theta(h_i^t)$. The output, $b_i\in R^{N\times{M}}$, predicts $M$ motion features of the $N$ agents. The specific motion features to be predicted depend on applications. The same decoder is used by all agents to decode their local states. Again a shared model will maximize the training effectiveness. 

We hypothesize that the more effectively an agent understands its environment, the better its decisions will be. A thorough understanding of the environment requires efficient feature extractions from  observations and messages, as well as accurate state updates that retain useful information and discard irrelevant history. Relying solely on deep reinforcement learning (DRL) to train the feature extractor and LSTM is insufficient, as the game's reward signal does not provide direct feedback on state representation. We propose that by training the agent to accurately predict the future motion and rewards of its teammates from its hidden state $h_i^t$, we can improve the feature extractor and LSTM, leading to a more refined representation of the hidden state.

Another advantage of these auxiliary prediction tasks is that they encourage the retention of teammate information within the agent's hidden state. Since this information may not directly influence the agent's reward, it could be overlooked if the system is trained solely using DRL. However, retaining such information is crucial for making collaborative decisions, such as communication. The auxiliary predictive tasks are only performed during the training. The belief decoder is not required during the testing.

\subsection{Loss Functions}
We employ centralized training and distributed execution, with all agents sharing the same BEPAL model and being trained together.

The training loss for each agent is composed of two main components: the actor-critic loss $L_{RL}$ and the auxiliary objectives losses $L_{aux}$, represented as $Loss =  L_{RL} + \lambda L_{aux}$, where $\lambda$ is a hyper-parameter. The auxiliary loss is derived by comparing the agents' beliefs with the ground truth, specifically $\overline{b^t}$ and $\overline{p^t}$. The Mean Squared Error (MSE) is utilized to quantify the auxiliary objectives loss, as outlined in the equation:
\begin{equation}
L_{aux}= \sum_t^T \sum_i^N \left(\mu MSE(\overline{b^t}-b_i^t) + \nu MSE(\overline{p^t}-p_i^t)\right)
\label{feedback_loss}
\end{equation}
where $N$ represents the number of agents, and $T$ denotes the total number of time steps. $\mu$ and $\nu$ are additional hyperparameters indicating which belief type (motion or reward) is more important.  During training, a central controller is needed to monitor all agent motion and observations and generates the ground truths. 

The actor-critic loss is $L_{RL} = L_{actor}+ \beta L_{critic}$, where $\beta$ is an importance hyper parameter. The critic loss aims to minimize the errors between the estimated state value and the discounted return:
\begin{equation}
L_{critic} = \sum\nolimits_{t} || r_i^t + \gamma V(h_i^{t+1}) - V(h_i^t) || ^2 
\label{critic_loss}
\end{equation}
where $r_i^t = \mathcal{R}(h_i^t, a_i^t)$ specifies the reward for agent $i$ at time $t$, and $V()$ denotes the value estimation by the critic model. The actor loss aims to maximize the expected cumulative reward by updating the policy parameters in the guided direction of critic. The actor network is updated via policy gradient as:
\begin{equation}
    \nabla_{\theta} J(\theta)  = \sum\nolimits_{t} \nabla_{\theta} \log \pi_{\theta}(a_i^t | h_i^t) [r_i^t + \gamma V(h_i^{t+1}) - V(h_i^{t})]
\end{equation}
where $\pi_{\theta}$, parameterized by $\theta$, represents the policy function. The BEPAL-MAS model is updated with the average gradient of all agents, aiming to achieve consistent improvements across the system. 

\section{Experiments}
\begin{figure*}[h]
  \centering
    \begin{subfigure}[b]{0.3\textwidth}
        \centering
        \includegraphics[scale=0.25, trim={1.5cm 0cm 1.5cm 0cm}]{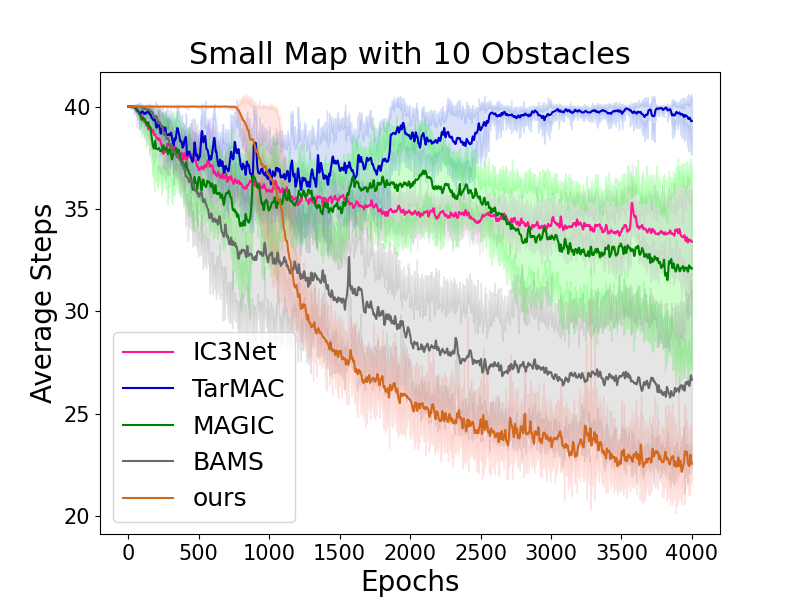}
        \caption{Small map, 5 agents, 10 obstacles}
        \label{fig:sub1}
    \end{subfigure}
    \hspace{5mm}
    \begin{subfigure}[b]{0.3\textwidth}
        \centering
        \includegraphics[scale=0.25, trim={1.5cm 0cm 1.5cm 0cm}]{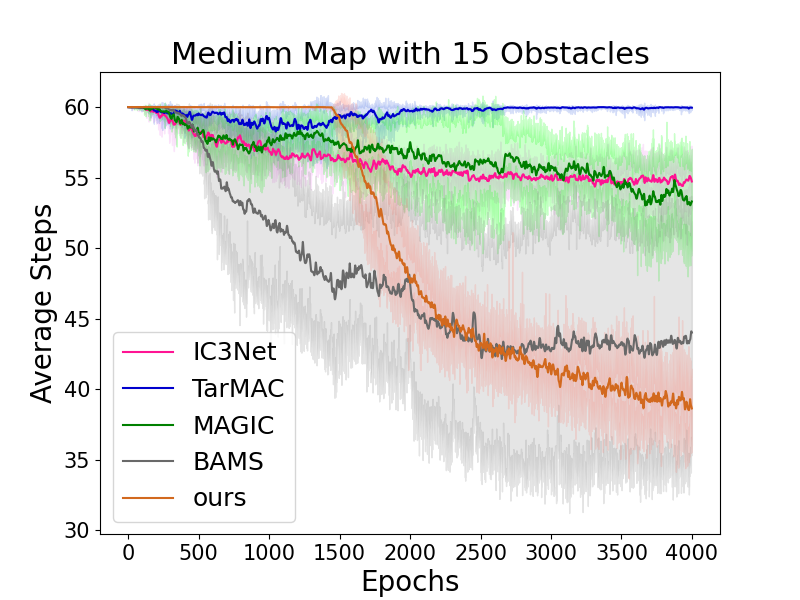}
        \caption{Medium map, 7 agents, 15 obstacles}
        \label{fig:sub2}
    \end{subfigure}
    \hspace{5mm}
    \begin{subfigure}[b]{0.3\textwidth}
        \centering
        \includegraphics[scale=0.25, trim={1.5cm 0cm 1.5cm 0cm}]{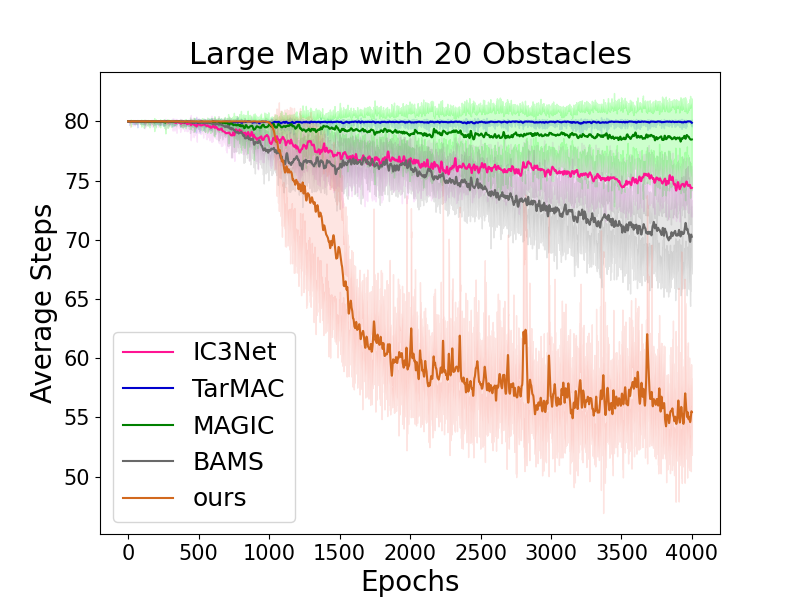}
        \caption{Large map, 5 agents, 20 obstacles}
        \label{fig:sub3}
    \end{subfigure}
    
    \caption{Performance comparison for the Predator-Prey games with obstacles is conducted across maps of different sizes: Small ($12\times{12}$), medium ($16\times{16}$) and large ($20\times{20}$)}
  \label{env}
  \vspace{-5pt}
\end{figure*}
\subsection{Evaluation Environments}
To evaluate the performance of the BEPAL-based multi-agent system (BEPAL-MAs), we conducted experiments in two distinct environments: Predator-Prey and the complex Google Research Football (GRF) environment.

\textbf{Predator Prey (PP)} is a widely used MARL benchmark where $N$ predators (agents) with a limited vision range ($3\times3$) search for a stationary prey in a grid-based environment. Both the agents and prey are randomly initialized in the environment. The agents receive a -0.05 penalty until they catch the prey, after which they no longer receive a penalty. The game concludes when all predators have reached the prey and the goal is to complete the game with the fewest steps. We tested environments with three complexity levels by varying the map size and the number of randomly placed obstacles that agents cannot pass through as shown in Fig. \ref{env}.

\textbf{Google Research Football (GRF)} 
is a physics-based 3D soccer simulator designed for reinforcement learning. GRF offers 19 actions (e.g., movement, kicking, dribbling). The field dimensions are  $2\times{0.82}$. We evaluated the models in the academy scenario 3 vs. 1 setting, where three attackers controlled by our model face off against one defender and one goalkeeper controlled by the built-in AI. To simulate a partially observable environment, we limited the agents' vision range to a radius of 0.5. The game concludes when a goal is scored, the ball goes out of bounds, or the possession changes. Each agent receives a reward of $+1$ for scoring a goal, 0 otherwise. Performance was measured by the average success rate, with a higher rate indicating superior model performance.

\subsection{Experiment setup}
\label{setup}
We optimized our network using RMSprop \cite{ruder2016overview}, with a learning rate of 0.001 and a smoothing constant of 0.97. Entropy regularization was applied with a coefficient of 0.01. Coefficient of critic loss $\beta$ is set to 0.05 and coefficient of auxiliary objective loss $\lambda=(\mu,\nu)$ is set to $(0.05/N, 0.5)$. The discounting factor for both games are set to be 1 since the outcomes of both games are given at the end. These hyperparameters were selected by hand and found to work well in practice, although a more systematic grid search could be used in future work. Code will be released upon publication.%The implementation of BEPAL is available at: https://anonymous.4open.science/r/BEPAL-D475/

We compare BEPAL with four baseline MARL models that leverages inter-agent communications with individual controllable communication gate: IC3Net, which employs message gating; TarMAC, which uses both message gating and attention; MAGIC, which leverages GNN-based models for message generation and passing in a centralized execution setup; and BAMS, which uses a grid-shaped belief map as one feedback channel. All experimental results are averaged over 5 seeds.

\subsection{Main Results for Predator Prey Game}
\subsubsection{Performance Comparison}

Table \ref{PP-main} compares the BEPAL with four baseline algorithms in terms of average number of steps required to complete a game in an environment without obstacles. In the small (i.e., $m=12$) and medium maps (i.e., $m=16$ ), BEPAL-MAS performs comparably to the best baseline in terms of average steps, however, with a lower standard deviation. On the larger map (i.e., $m=20$), it achieves a 18.62\% reduction in average steps and a 50\% reduction in the standard deviation compared to the best baseline.

\begin{table}[t]
\vspace*{-3pt}
\captionsetup{justification=centering}
  \caption{Average steps comparison in predator prey environment without obstacle}
  \label{PP-main}
  \centering
    {\footnotesize
    \begin{tabular}{lccc} 
    \toprule
     &{N=5 m=12}        &{N=7, m=16}        &{N=5, m=20}   \\
     & Max Steps = 40   & Max Steps = 60     & Max Steps = 80  \\
    \midrule
    IC3Net       & 32.90 $\pm$ 5.22 & 53.54 $\pm$ 7.51 & 71.34 $\pm$ 8.16 \\
    TarMAC       & 34.59 $\pm$ 10.48 & 44.68 $\pm$ 9.19  & 79.94 $\pm$ 5.98 \\
    MAGIC        & 26.89 $\pm$ 9.73 & 46.04 $\pm$ 17.94 & 78.36 $\pm$ 9.49 \\
    BAMS         & 21.64 $\pm$ 4.26 & 30.76 $\pm$ 6.98 & 53.79 $\pm$ 16.49  \\
    Ours    & \textbf{18.23 $\pm$ 2.41}  & \textbf{29.38 $\pm$ 5.42} & \textbf{43.77 $\pm$ 8.48}     \\
    \bottomrule
    \end{tabular}
    }
\end{table}
We then introduced obstacles to create  a more complex environment. Figure \ref{env} shows the average number of steps required for all 5 models to complete the game. Compared to the baselines, our method demonstrated  superior performance, reducing average step counts by 15.42\%, 10.82\% and 21.67\%  in the small, medium and large maps, compared to the best baseline model, respectively (Figure \ref{fig:sub1}-\ref{fig:sub3}). Moreover, our method achieved more stable learning, with a 60.91\% and 73.80\% reduction in standard error compared to the best baseline.

Notably, our method exhibits slower initial learning during the early epochs due to the additional loss introduced by auxiliary learning, which must be optimized concurrently. As a result, the learning process progresses more slowly at the beginning. However, once the auxiliary tasks are properly learned, our method surpasses all baselines in learning speed. This highlights the effectiveness of auxiliary learning in facilitating policy learning and improving overall performance.

It’s important to point out that, although the standard deviation of the baselines decreases on the large map, this does mean their performance became more robust. Instead, this is due to the fact that agents using the baseline model often fail to complete the game within the maximum allowable steps. As a result, the number of steps taken remains to be the maximum value in most games. 

\subsubsection{Ablation Study}To quantify the impact of the auxiliary learning, we conducted an ablation study in an environment with a $12\times{12}$ map, 5 agents and 10 obstacles. Table \ref{ablation} shows that after removing both reward and motion predictions, the number of steps taken by BEPAL-MAS increased 77.2\%. When only the reward prediction or motion prediction was removed, the average number of steps increased by 10.4\% and 66.0\% respectively. The results support our hypothesis that performing auxiliary learning helps agents develop a robust understanding of the environment and achieves a better hidden state representations, and consequently, lead to a better game performance.

\begin{table}[t]
  \centering
  \caption{Ablation study on BEPAL}
  \label{ablation}
  \centering
  {\begin{tabular}{lcc}
    \toprule
    Environment     & GRF & PP with Obstacle      \\
    \midrule
    Evaluation Metric    & Success Rate & Average Steps      \\
    \midrule
    BEPAL                           & 83.64\%   & 22.21  \\
    BEPAL w/o Reward Prediction     & 72.97\%   & 24.52  \\
    BEPAL w/o Motion Prediction     & 19.14\%   & 36.86  \\
    BEPAL w/o Auxiliary Learning     & 8.35\%   & 39.37  \\
    \bottomrule
  \end{tabular}}
\end{table}

\begin{table}[t]
  \captionsetup{justification=centering}
  \vspace{5pt}
  \caption{Transferability study in PP Environment: Comparing native and transfer models for number of steps to complete the game}
  \label{PP-scala}
  \centering
    \begin{tabular}{lccc}
    \toprule
     &{N=5 m=12}    &{N=5, m=12}  &{N=5, m=20}  \\
     & Max Steps = 40 & Max Steps = 40  & Max Steps = 80  \\
     & obstacles = 0 & obstacles = 10  & obstacles = 20  \\
    \midrule
    Native       & 18.23  & 22.41  & 55.23     \\
    Transfer         & 19.76  & 22.41   & 54.62    \\
    
    \bottomrule
    \end{tabular}
\end{table}
\subsubsection{Transferability}To evaluate the transferability of BEPAL, we trained it in an environment with $12\times{12}$ map, 5 agents and 10 obstacles. The trained model is then transferred to different game settings and its performance is tested. As a comparison, we also trained a native model using the configuration of the testing environment. The performance of the "Transferred" and "Native" models are compared in Table \ref{PP-scala}. The results show that BEPAL has a robust performance and can be transferred to different testing environments.

It is interesting to note that, for large map (i.e., $m=20$), the transfer model works better than the native model. This is probably because, given a more complex environment, the native trained model requires much longer training time. Hence, under the same number of training epoch, the transfer model performs better, which is an example of curriculum learning \cite{bengio2009curriculum}. 

\subsubsection{Correlation between Auxiliary and Reinforcement Learning}\label{corr_explain} 
\begin{figure}[t]
\vspace{3pt}
  \centering
    \begin{subfigure}{0.48\columnwidth}
        \centering
        \includegraphics[scale=0.2, width=\textwidth]{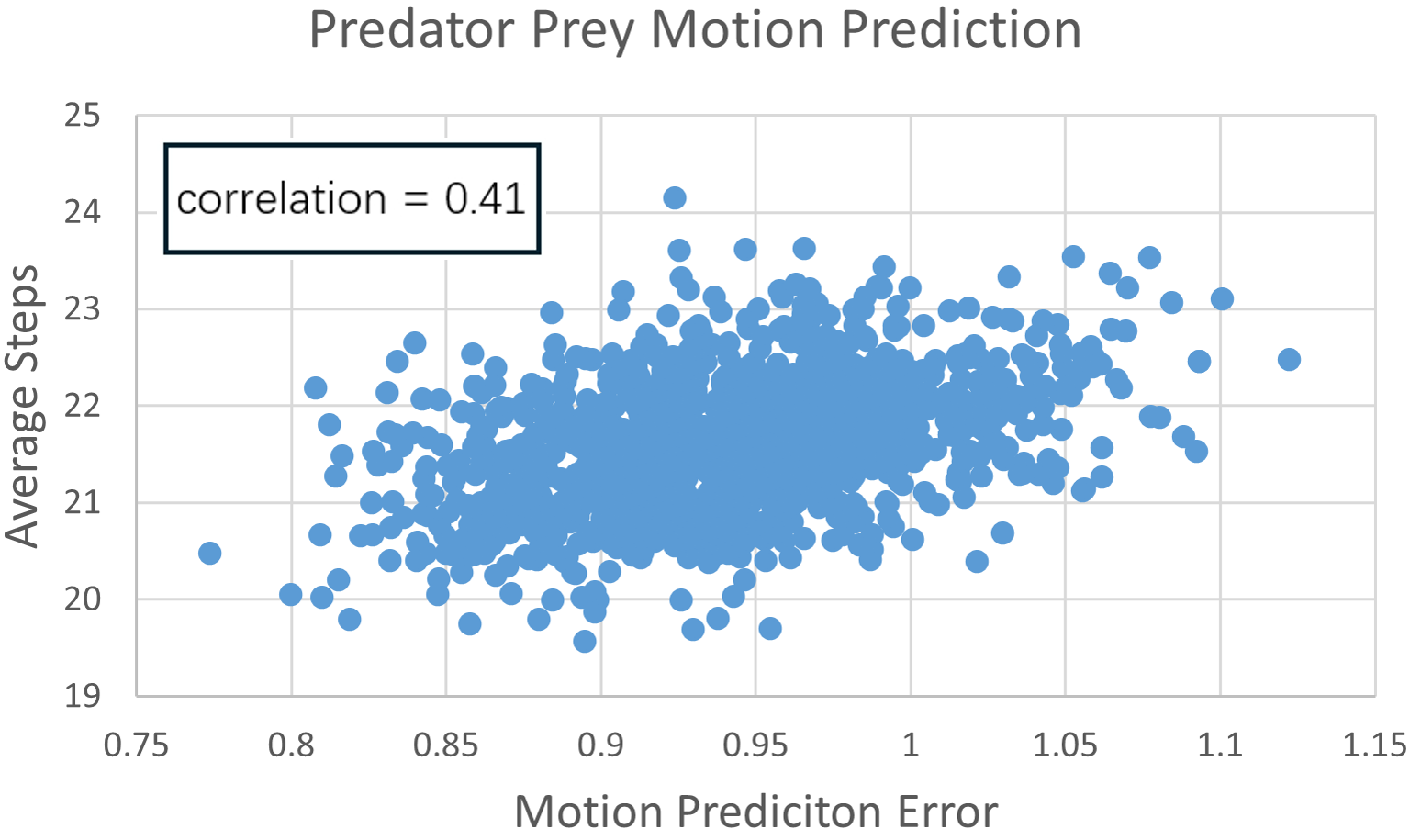}
        \caption{PP Motion Prediction}
        \label{motion}
    \end{subfigure}
    \hfill
    \begin{subfigure}{0.48\columnwidth}
        \centering
        \includegraphics[scale=0.2, width=\textwidth]{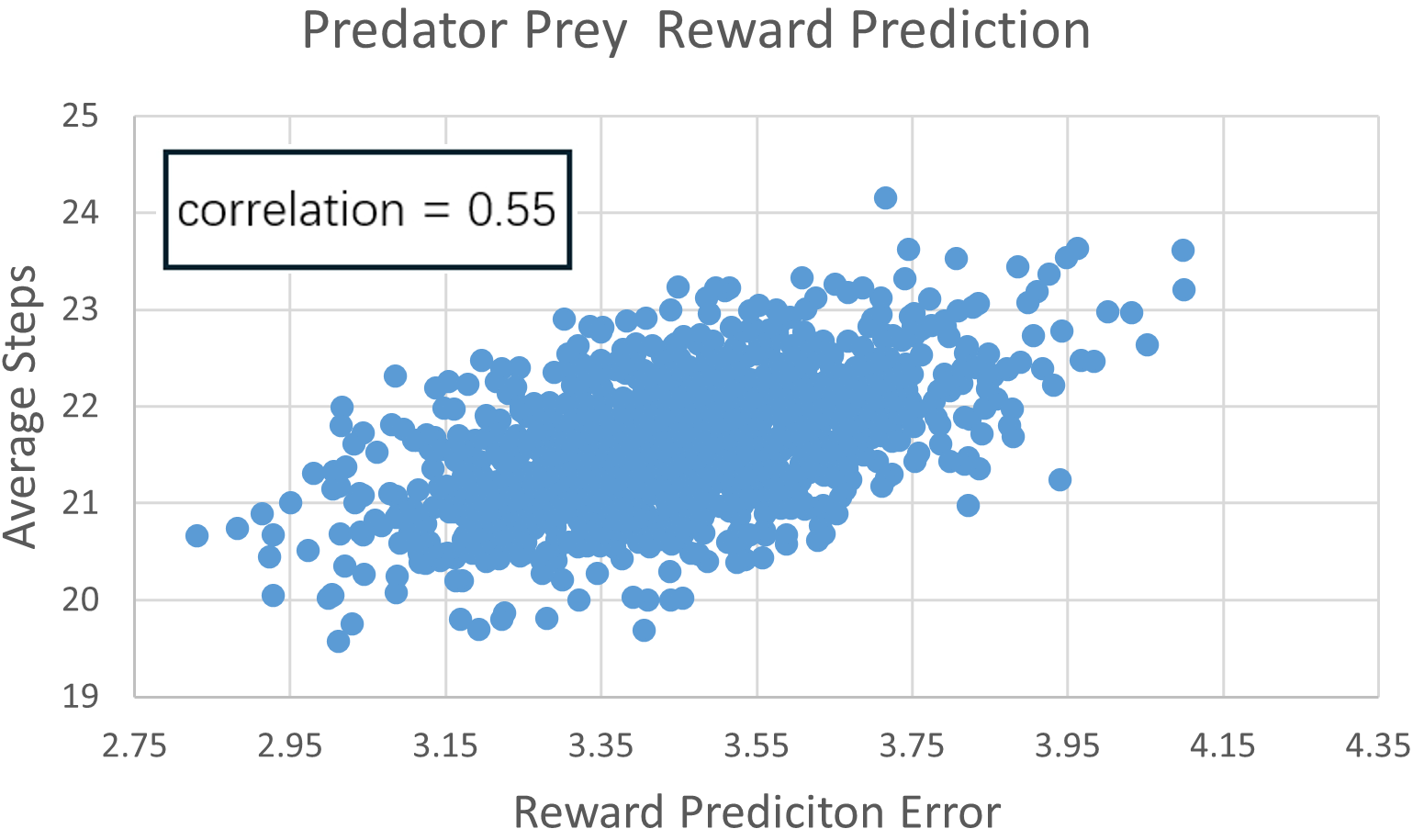}
        \caption{PP Reward Prediction}
        \label{reward}
    \end{subfigure}
    \hfill
    \begin{subfigure}{0.48\columnwidth}
        \centering
        \includegraphics[scale=0.2, width=\textwidth]{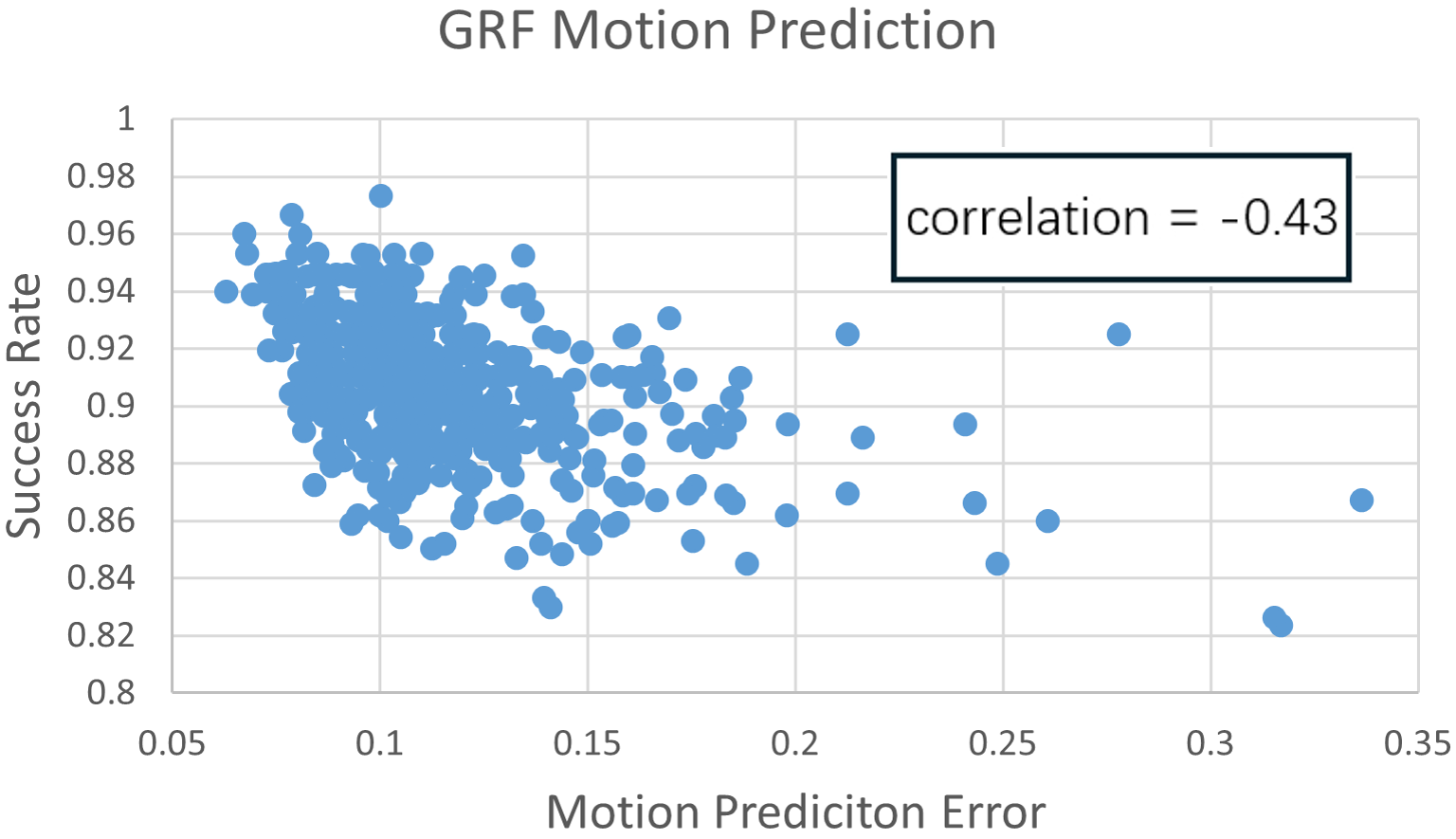}
        \caption{GRF Motion Prediction}
        \label{motion}
    \end{subfigure}
    \hfill
    \begin{subfigure}{0.48\columnwidth}
        \centering
        \includegraphics[scale=0.2, width=\textwidth]{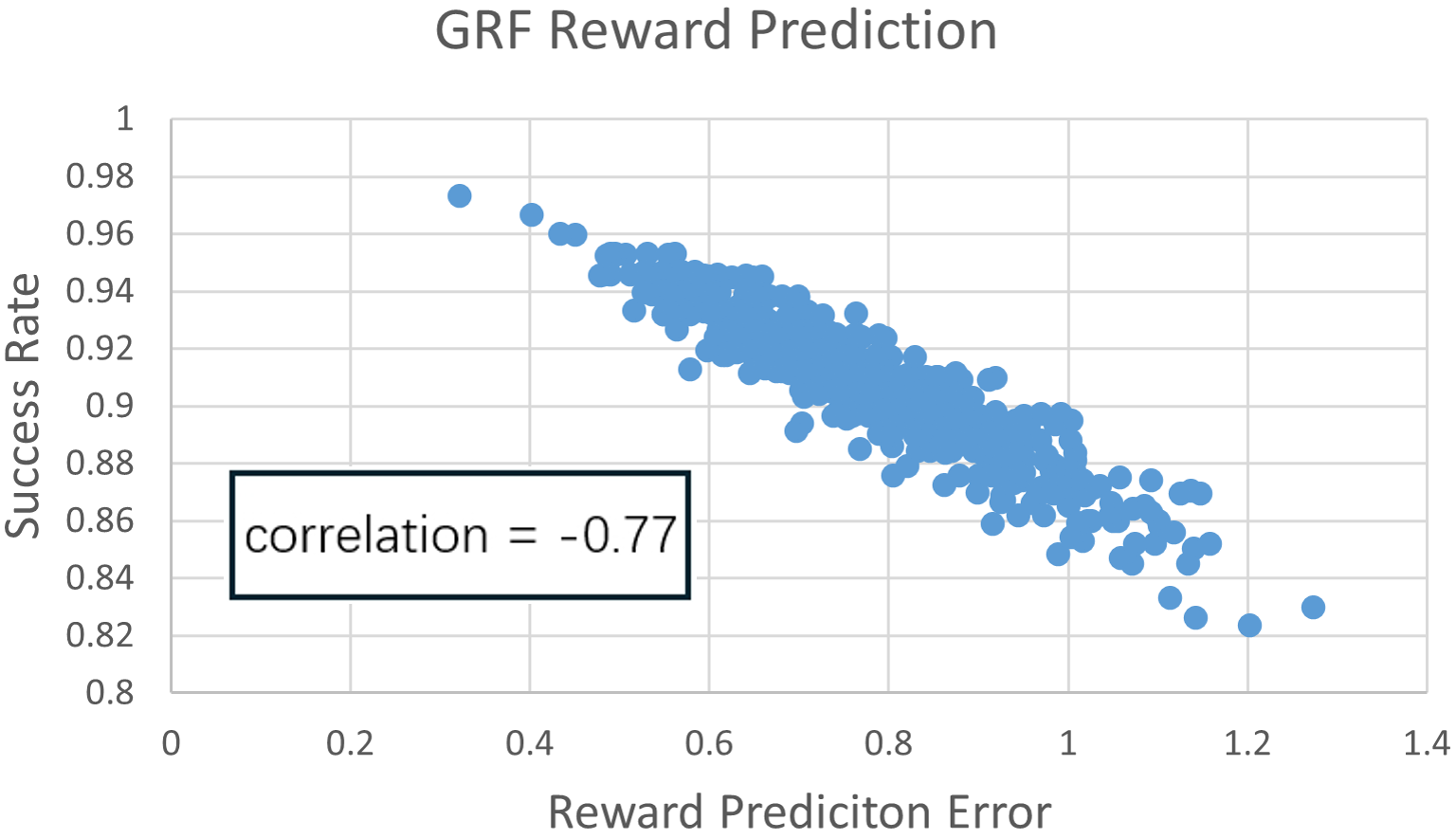}
        \caption{GRF Reward Prediction}
        \label{reward}
    \end{subfigure}
    \caption{Game Performance vs. Auxiliary Prediction Accuracy.}
  \label{aux_pre}
  %\vspace{-7pt}
\end{figure}

Experiments were conducted to demonstrate the correlation between the auxiliary learning and the MARL learning. Figure \ref{aux_pre} illustrates the relationship between the accuracy of auxiliary prediction and the number of steps agents needed to complete the game. On average, the agent's performance in the game shows an average Pearson correlation coefficient of 0.41 with the accuracy of motion prediction and 0.55 with the accuracy of reward prediction on two games, both with a p-value close to 0. This positive correlation suggests that training the agent to perform auxiliary predictions enhances its performance in the game.

While the accuracy of reward prediction shows a higher correlation with agent performance than that of motion prediction, Table \ref{ablation} shows that it has a comparatively smaller influence on improving agent performance. This can be attributed to the intrinsic relationship between reward prediction and the critic model. Since both the critic and the reward prediction model are responsible for predicting rewards, albeit for different scopes (individual versus team), the performance of these two models are strongly correlated. Meanwhile, the quality of the critic model determines the success rate of the agent performance. Hence the accuracy of the reward predictor is strongly correlated to the game performance of the agent. Higher correlation with the agent performance does not mean the reward prediction provide more information to improve the RL training. On the contrary, because its high correlation with the the critic model, reward prediction does not have as significant impact to the RL training as motion prediction. %However, this does not imply that motion prediction is less influential in enhancing performance. Motion prediction contributes critical information about agent dynamics, which complements reward prediction and supports overall decision-making. Together, the two auxiliary tasks provide distinct but synergistic benefits to the agent's learning process.

\subsubsection{Computation Overhead of Auxiliary Learning} To evaluate whether the auxiliary tasks introduce significant computational overhead, we conducted experiments comparing the training times for models with and without auxiliary tasks. The experiments were performed on the same Intel(R) Core(TM) i7-8700K CPU @ 3.70GHz, with the average time per epoch calculated over 100 epochs. The BEPAL model with auxiliary tasks required an average of 73.98 seconds per epoch, while the model without auxiliary tasks took 68.67 seconds per epoch. This result demonstrates that incorporating auxiliary learning increases computation time by only 7.82\% per epoch, while providing an average 16\% improvement in performance. A learned auxiliary predictor also helps the RL training to converge faster, and eventually requires fewer epochs. These findings confirm that the addition of auxiliary tasks is computationally efficient and significantly beneficial to overall performance.

\begin{figure*}[htbp]
  \centering
  \begin{subfigure}[b]{0.9\textwidth}
        \centering
        \includegraphics[scale=0.5]{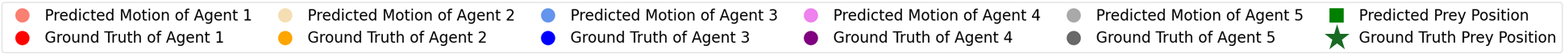}
        \label{fig2:sub6}
    \end{subfigure}
    \begin{subfigure}[b]{0.2\textwidth}
        \centering
        \includegraphics[scale=0.2, trim={1.5cm 0cm 1.5cm 0cm}]{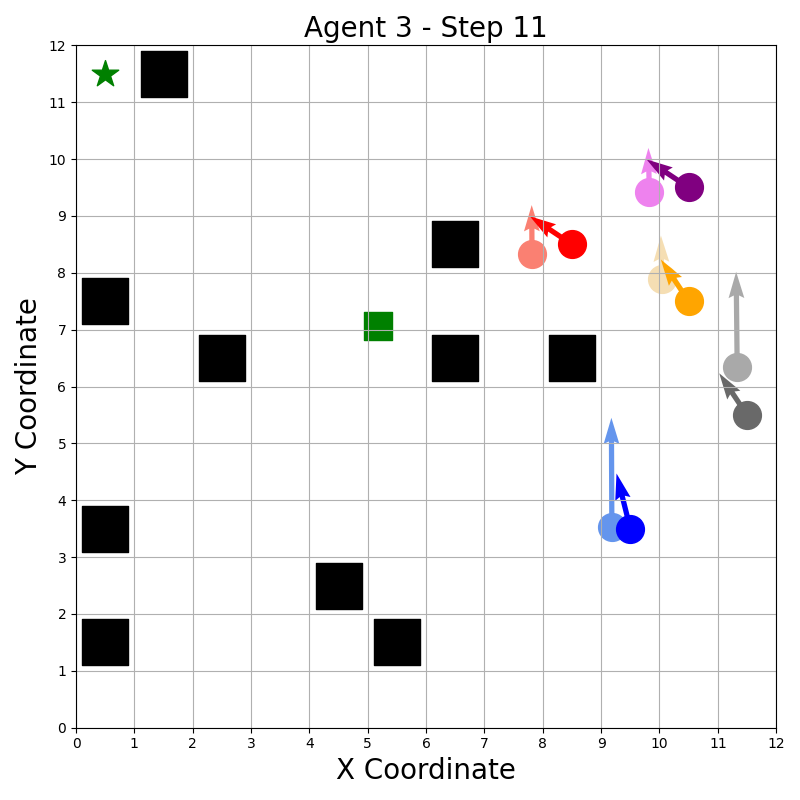}
        \caption{Agent 3's Prediction}
        \label{fig2:sub1}
    \end{subfigure}
    \hspace{5mm}
    \begin{subfigure}[b]{0.2\textwidth}
        \centering
        \includegraphics[scale=0.2, trim={1.5cm 0cm 1.5cm 0cm}]{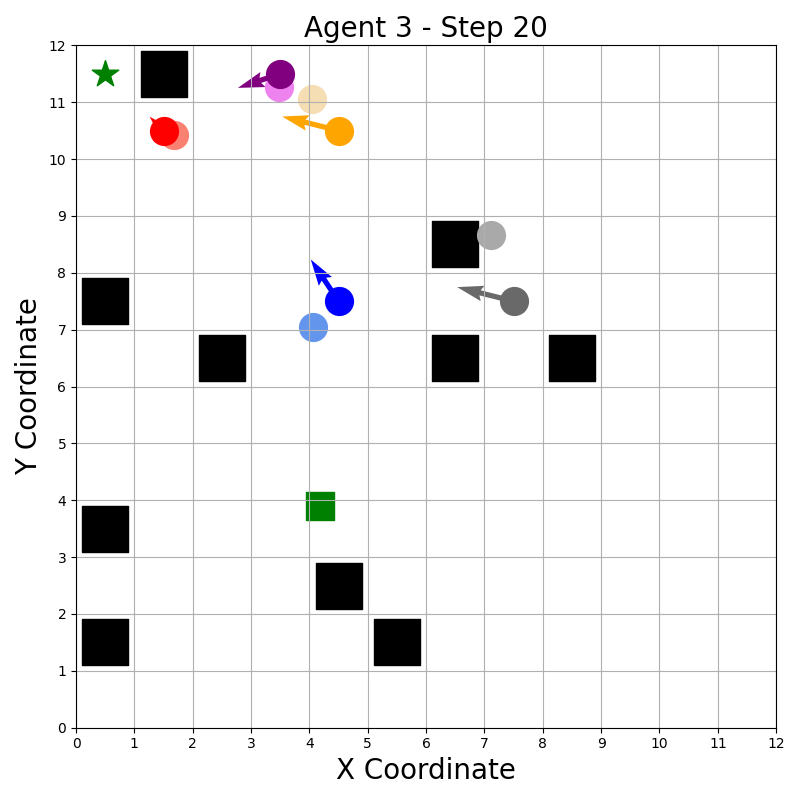}
        \caption{Agent 3's Prediction}
        \label{fig2:sub2}
    \end{subfigure}
    \hspace{5mm}
    \begin{subfigure}[b]{0.2\textwidth}
        \centering
        \includegraphics[scale=0.2, trim={1.5cm 0cm 1.5cm 0cm}]{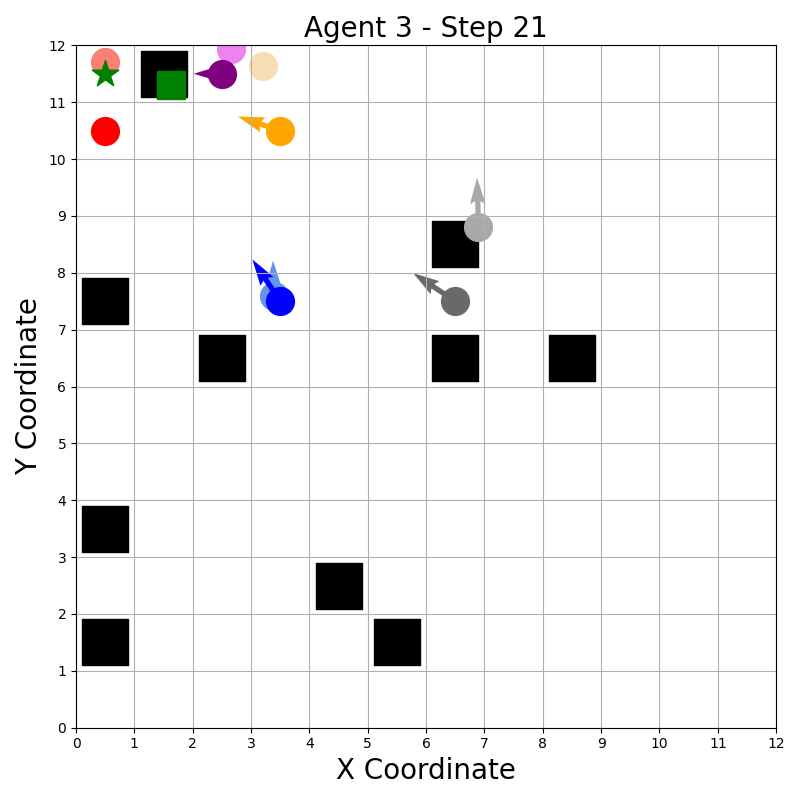}
        \caption{Agent 3's Prediction}
        \label{fig2:sub3}
    \end{subfigure}
    \hspace{5mm}
    \begin{subfigure}[b]{0.2\textwidth}
        \centering
        \includegraphics[scale=0.2, trim={1.5cm 0cm 1.5cm 0cm}]{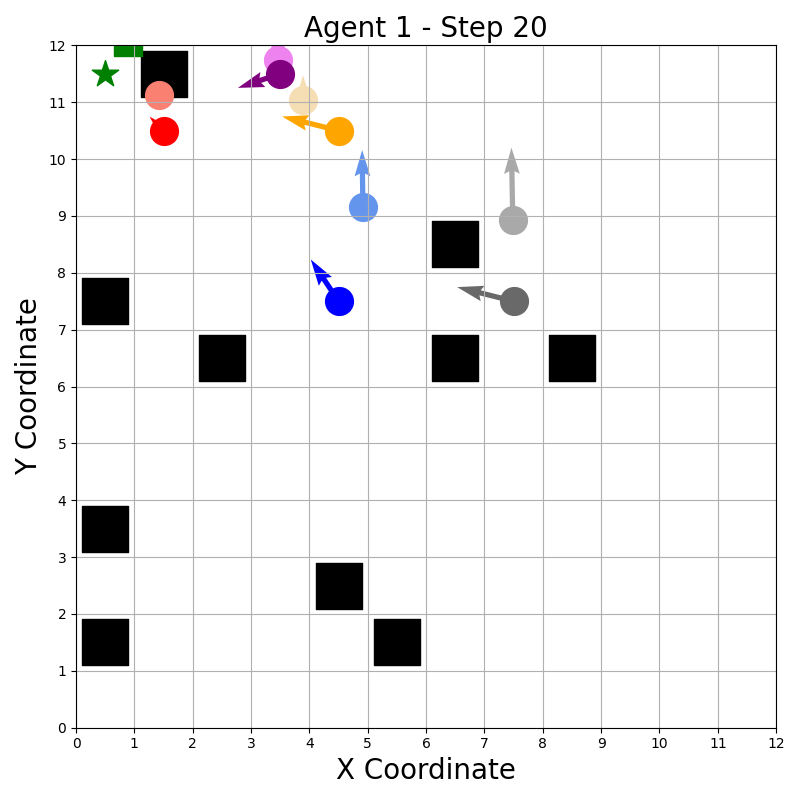}
        \caption{Agent 1's Prediction}
        \label{fig2:sub4}
    \end{subfigure}
    
    \caption{Visualization of motion prediction generated by agents in different time steps in a $12\times{12}$ map with 10 obstacles. (a-c) Predictions generated by agent 3 at time steps 11, 20, and 21. (d) Prediction generated by agent 1 in time step 20. }
  \label{visual}
  \vspace{-5pt}
\end{figure*}

\subsubsection{Visualization of Auxiliary Predictions}  We visualize some motion predictions captured from a PP game. Fig. \ref{visual}a-c show the motion prediction made by agent 3 in steps 11, 20, and 21; and Fig. \ref{fig2:sub4} shows the prediction made by agent 1 in step 20. The prediction includes the coordinates and moving directions  (represented as vectors) of all agents, including the predictor itself, in the next several time steps and the location of the prey. The prediction represents the agent's interpretation of the current environment state.

As shown in Fig. \ref{fig2:sub1}, at time step 11 when no agent has discovered the prey, agent 3 predicts that the prey location is in an unexplored area to motivate exploration. Its prediction of other agent's motion is a mixture of information from the received messages and its own prediction of the prey location. At time step 20, agent 1 observed the prey (Fig. \ref{fig2:sub4}) and broadcast the information. Meanwhile, as shown in Fig. \ref{fig2:sub2}, agent 3 and other agents have not yet received the message, they move from right side of the map, which has been explored, to the left side, which is unexplored. Finally, in step 21 (Fig. \ref{fig2:sub3}), all agents received the message from agent 1, and they all move toward the true location of the prey. As we can see the predicted motion generally aligns with the actual motion of the agents.  Without receiving new messages, the predicted movement distance (arrow length) for each agent (Fig. \ref{fig2:sub2}) reduces as the agent is no long sure about the motion of its teammates. Once the agent acquires critical information, it starts predicting larger movement distances (Fig. \ref{fig2:sub3}), reflecting increased certainty in its updated information.

\begin{figure}[t]
\vspace{-3pt}
\centering
\includegraphics[scale=0.3]{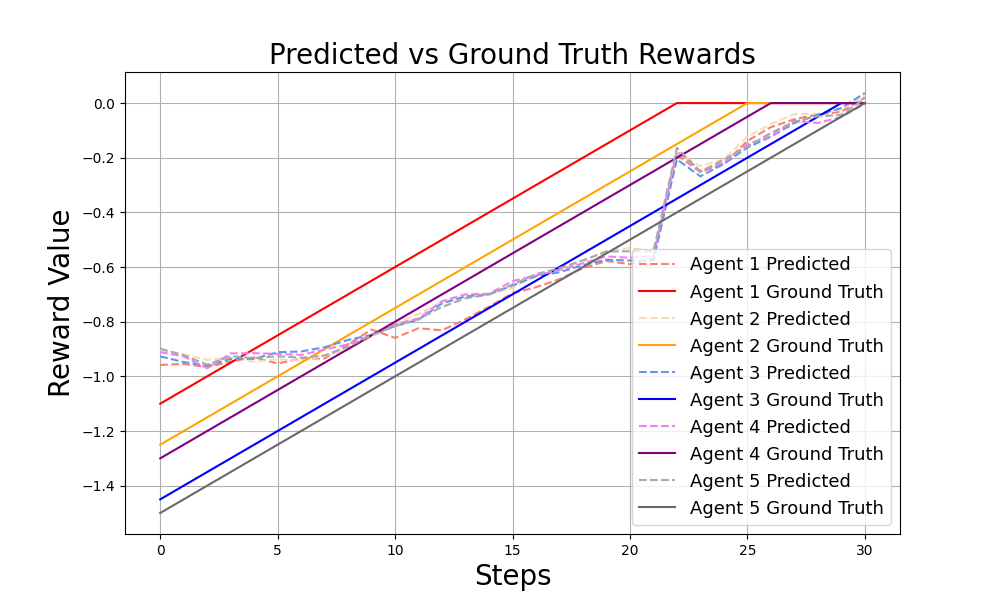}
  \caption{Visualized Reward Prediction}
  \label{reward_prediction}
  \vspace{-3pt}
\end{figure}

Fig. \ref{reward_prediction} illustrates the reward predictions generated by agent $3$ in the previous game. In the early steps of the game, the agent predicts future rewards with a negative average value, reflecting its learned expectation. As exploration progresses, the predicted rewards gradually increase, indicating steady progress in exploration. A sudden change in reward predictions signifies that the prey has been located. Subsequently, the agent updates its predictions of other agents' rewards, reflecting the belief that the game will soon end.

\subsection{Main Results for Google Research Football}

\begin{figure}[t]
\vspace{3pt}
\centering
\includegraphics[scale=0.25]{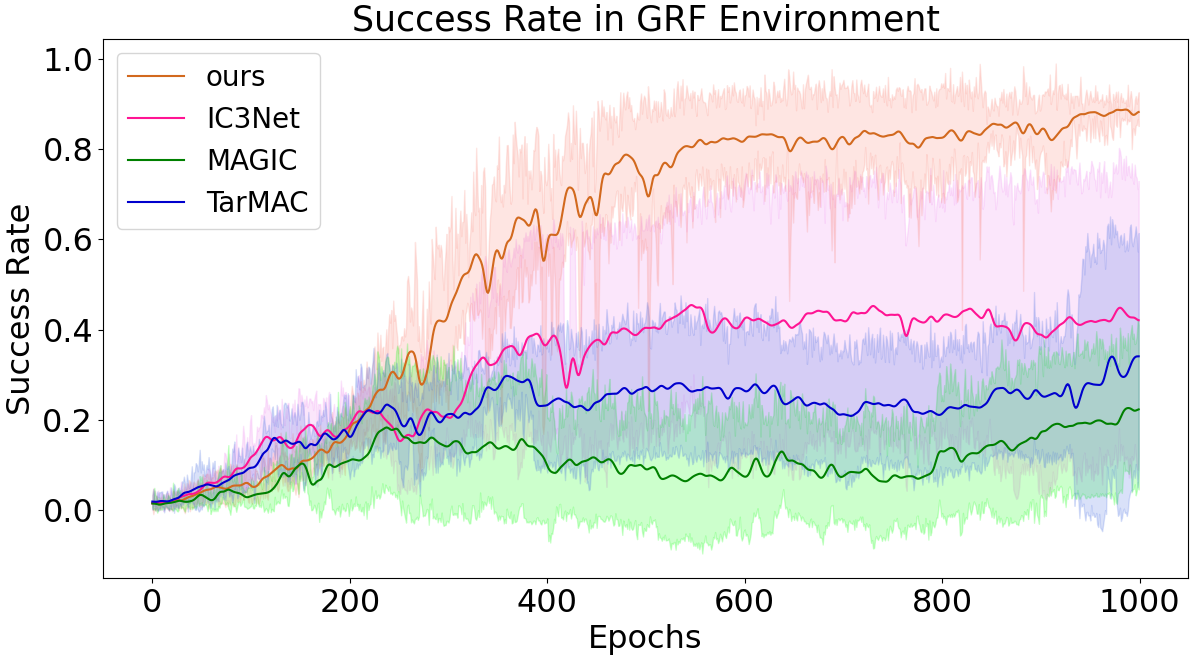}
  \caption{Comparison of baselines on GRF environment}
  \label{grf_res}
  
\end{figure}
The GRF environment features stochastic state transitions, sparse rewards, and a complex action space. The evaluation metric is the scoring success rate. We modified the GRF, originally a fully observable environment, into a partially observable one by masking objects outside the agent's visual range. It is important to note that after this modification, the performance of MAGIC and TARMAC declined significantly compared to previously reported results \cite{niu2021multi}.

Figure \ref{grf_res} compares the change in the success rate over training epochs between BEPAL and three baseline models. Note that BAMS could not be applied to GRF due to its CNN-based observation encoder being incompatible with the non-grid environment of GRF. The results show that BEPAL converges to a higher success rate at a faster speed with more stable learning compared to all baselines.

 Table \ref{ablation} presents the results of the ablation study, demonstrating the impact of the two auxiliary learning tasks on the GRF game. As shown in the table, omitting both auxiliary tasks results in a 90\% drop in the success rate. The success rate decreases by 12.7\% and 77.1\% when the reward prediction or motion prediction tasks are omitted, respectively. These results highlight the effectiveness of auxiliary learning in enhancing agent game performance.

The relationship between the prediction error and game performance is also shown in Figure \ref{aux_pre}. The effectiveness of both predictions follows the same trend as observed in the PP environment. However, we observe that GRF exhibits a particularly strong correlation between reward prediction and success rate. This is because GRF provides a global reward shared by all agents. The more accurate an agent can predict team member's future reward, the better it can predict its own expected reward, which consequently lead to better action.
%leading to uniform reward values across the team. Consequently, this results in a stronger correlation between reward prediction and success rate compared to the PP environment. 
However, as discussed in Section \ref{corr_explain}, a stronger correlation with reward prediction does not necessarily indicate a greater influence on final performance.  %We observe that, for both PP and GRF games, including motion prediction provides more benefits than including reward prediction. However, as shown in Figure \ref{aux_pre}, the accuracy of reward prediction has a stronger correlation with game performance. This is because reward prediction also includes the agent's own reward, which aligns with its own value function. While the agent's value function highly correlates to the game performance, it does not provide additional information to the RL process. 

\section{Conclusions}
This paper proposes a novel training approach that uses auxiliary learning to enhance multi-agent reinforcement learning in a partially observable environment. The auxiliary objective functions provide additional feedback during training, refining the agent’s hidden state representation and enriching it with information that, while not directly contributing to the agent’s reward, benefits in-agent collaboration. The auxiliary learning leverages a belief decoder to predict its teammates’ motion and reward information from the agent’s hidden state. By training on these auxiliary prediction tasks, the agent gains a better understanding of its environment and becomes more aware of its teammates' status.

We compared our method with IC3Net, MAGIC, and TarMAC in both predator-prey environments (with and without obstacles) and the Google Research Football (GRF) environment. Our approach demonstrates that auxiliary learning significantly improves the agent’s game performance and makes the training process more robust.

% \section*{Acknowledgment}

% This research is partially supported by the Air Force Office of Scientific Research (AFOSR), under contract FA9550-24-1-0078 and NSF under award CNS-2148253. Approved for Public Release; Distribution Unlimited PA Clearance Pending- Do Not Distribute.

\nocite{*} 
\bibliographystyle{ieeetr}
\bibliography{conference_101719}

\end{document}